\documentclass[10pt,a4paper]{article}

\usepackage{amsmath}
\usepackage[psamsfonts]{amssymb}
\usepackage{latexsym}
\usepackage{cmmib57}
\usepackage{pstricks}
\usepackage{pst-coil}
\usepackage{pst-node}

\newcommand{\scr}[1]{\mathcal{#1}}
\def\idty{{\leavevmode{\rm 1\ifmmode\mkern -4.8mu\else\kern -.3em\fi
      I}}}
\renewcommand{\Bbb}[1]{\if1#1\idty\else\mathbb{#1}\fi}
\renewcommand{\tilde}{\widetilde}

\newcommand{\kb}[1]{|#1\rangle\langle#1|}

\newcommand{\BK}[2]{\langle#1|#2\rangle}
\newcommand{\ket}[1]{|#1\rangle}

\newcommand{\tr}{\operatorname{tr}}
\newcommand{\diag}{\operatorname{diag}}
\newcommand{\Id}{\operatorname{Id}}

\newcommand{\Al}{\mathcal{A}}
\newcommand{\Bl}{\mathcal{B}}
\newcommand{\Pl}{\mathcal{M}}

\newcommand{\draw}{\emptyset}
\newcommand{\hd}[1]{\noindent\textit{#1}}

\newgray{meascolor}{.5}
\newgray{opercolor}{.5}
\newgray{entacolor}{.5}
\newgray{qbitcolor}{.1}
\newgray{mklight}{.7}
\newgray{mkfill}{.9}
\psset{unit=\unitlength}
\psset{arrowsize=2pt 5}
\psset{coilarm=.4}

\newtheorem{thm}{Theorem}[section]
\newtheorem{defi}[thm]{Definition}
\newtheorem{prop}[thm]{Proposition}

\newenvironment{proof}{\par\noindent\textit{Proof.\ }}{\hfill $\Box$ \vspace{1em}}

\begin{document}


\title{An introduction to quantum coin-tossing}
\author{C. D{\"o}scher\thanks{e-mail: \texttt{claus.doescher@t-online.de}}\quad and
  M. Keyl\thanks{e-mail: \texttt{m.keyl@tu-bs.de}}
  \\[1ex]
  {\small Institut f{\"u}r Mathematische Physik, TU Braunschweig,}\\
  {\small Mendelssohnstr.3, 38106 Braunschweig, Germany.}}
\maketitle


\begin{abstract}
  We review the quantum version of a well known problem of cryptography called coin tossing (``flipping a
  coin via telephone''). It can be regarded as a game where two remote players (who distrust each other)
  tries to generate a uniformly distributed random bit which is common to both parties. The only
  resource they can use to perform this task is a classical or quantum communication channel. In this
  paper we provide a general overview over such coin tossing protocols, concerning in particular their
  security.
\end{abstract}

\section{Introduction}

Coin flipping was introduced in 1981 by Blum \cite{Blum81} as a solution to the following cryptograhic
problem (cited from \cite{Blum81}): ``Alice and Bob want to flip a coin by telephone. (They have just
divorced, live in different cities, want to decide who gets the car.) Bob would not like to tell Alice
\emph{heads} and hear Alice (at the other end of the line) say: Here it goes... I'm flipping the
coin... You lost!''. Hence the basic difficulties are: both players (Alice and Bob) distrust each other,
there is no trustworthy third person available and the only resource they can use is the communication 
channel. Although this problem sounds somewhat artificial, coin tossing is a relevant building block 
which appears in many cryptograhic protocols. 

Within classical cryptography coin tossing protocols are in general based on assumptions about the
complexity of certain computational tasks like factoring of large integers, which are unproven and, even
worse, break down if quantum computers become available. A subset of classical cryptography which suffer
from similar problems are public key cryptosystems. In this case however a solution is available in form
of quantum key distribution (cf. \cite{QCRev} for a review) whose security is based only on the laws of
quantum mechanics and no other assumptions. Hence the natural question is: Does quantum mechanics provide
the same service for coin-tossing, i.e. is there a perfectly secure quantum coin-tossing protocol?
Although the answer is, as we will see, ``no'' \cite{LoCh98,MaSaChi}, quantum coin-tossing provides a
reasonable security improvement over classical schemes.

The purpose of this paper is to give an overview over this field, emphasizing in particular the game
theoretic aspects, and to review some recent results. Therefore its outline is as follows: In Section
\ref{sec:syst-stat-oper} we give a short survey on classical and quantum systems and operations on them.
This enables us in Section \ref{sec:coin-toss-prot} to develop a general scheme which allows the
description (and comparison) of quantum as well as classical coin tossing protocols (which are considered
in Section \ref{sec:class-coin-toss}), and which points out the game theoretic aspects of the subject. In
Section \ref{sec:unitary-normal-form} we show how many questions, in particular optimality, can be
reduced to a simplified scheme where only unitary operators and von Neumann measurements are involved. A
recent example is given in Section \ref{sec:particular-example} and some conclusions are drawn in Section
\ref{sec:bounds}. 

\section{Systems, states and operations}
\label{sec:syst-stat-oper}

In general a quantum protocol requires manipulation and exchange of quantum as well as classical
data. It is therefore useful, to have a unified description for all possible types of systems and
operations which we will encounter (this is only a brief survey; for a more detailed and complete
presentation see \cite{PRepQI}, Ch. 2 and 3.).  

\begin{itemize}
\item 
\hd{Quantum Systems:}  According to the rules of quantum mechanics, every kind of quantum systems is
  associated with a Hilbert space $\scr{H}$, which for the purpose of this article we can take as finite
  dimensional. The simplest quantum system has a two dimensional Hilbert space $\scr{H}=\Bbb{C}^2$ , and
  is called a \emph{qubit}, for `quantum bit'. The observables of the system are given by (bounded)
  operators. This space will be denoted by $\scr{B}(\scr{H})$. The preparations (states) are given by
  density operators, i.e. positive (trace-class) operators $\rho\in\scr{B}(\scr{H})$ with trace one.
\item 
\hd{Classical probability:} The classical analog of a state of a (finite dimensional) quantum system is
  a probability distribution $p_x, x \in X$ on a finite set $X$ of ``elementary events'', i.e. $X$
  describes the possible outcomes of a (classical) statistical experiment, like tossing a coin ($X =
  \{$head, number$\}$ or throwing a dice ($X = \{1,\ldots,6\}$), and $p_x$ is the probability that the outcome $x$
  occurs. Without loss of generality we will assume in the following that $X = \{1,\ldots,n\}$, $n \in \Bbb{N}$
  holds. The classical information contained in $p$ can be transformed easily into quantum information:
  We just have to prepare for each elementary event $x \in X$ an $n$-level quantum system (described by a
  Hilbert space $\scr{K}$) in a pure state $\kb{x} \in \scr{B}(\scr{K})$, where $\ket{x} \in \scr{K}$, $x \in
  X$ denotes a distinguished orthonormal basis and $\kb{x}$ is the projector onto $\ket{x}$. If the event
  $x \in X$ occurs with probability $p_x$, we get in this way quantum systems in the (mixed) state $\rho_p =
  \sum_x p_x \kb{x}$. Each state $\rho$ which is diagonal in the basis $\ket{x}$, i.e. $\rho = \sum_x \rho_x \kb{x}$,
  can be realized in this way, provided the initial probability distribution is $\rho_x, x \in X$. Now we
  introduce the space $\scr{C}(X) \subset \scr{B}(\scr{K})$ of \emph{diagoal} (with respect to the basis
  $\ket{x}$) operators on $\scr{K}$. According to our previous discussion we can identify the classical
  state space with the set of density operators in $\scr{C}(X)$.
\item 
\hd{Hybrid systems:} This point of view is very handy, if we want to describe a ``hybrid system'' which
  contains a classical part, described by the set $X$ and a quantum part, described by the Hilbert space
  $\scr{H}$: A state of a composite \emph{quantum} system, consisting of two subsystems with Hilbert
  spaces $\scr{H}$ and $\scr{K}$, is given by a density operator $\rho$ on the tensor product $\scr{H} \otimes
  \scr{K}$, i.e. $\rho \in   \scr{B}(\scr{H}) \otimes \scr{B}(\scr{K})$. 
  If one of the subsystem is classical, we only have to replace $\scr{B}(\scr{K})$ by $\scr{C}(X)$. Hence:
  states of a hybrid system can be described by density operators $\rho \in \scr{B}(\scr{H}) \otimes
  \scr{C}(X)$. It is easy to see that the elements of $\scr{B}(\scr{H}) \otimes \scr{C}(X)$ are operators which
  are block-diagonal of the form $\rho = \diag(\rho_1,\ldots,\rho_n)$ with $\rho_x \in \scr{B}(\scr{H})$.
\end{itemize}

Summarizing our discussion up to now we can say that all three kinds of systems can be described in terms
of a Hilbert space $\scr{H}'$ and a linear subspace $\scr{A} \subset \scr{B}(\scr{H}')$ which we will call in the
following the \emph{observable algebra}\footnote{This name originates from the fact that 1. $\scr{A}$ is
  in all three cases not only a linear space but a \emph{*-algebra} (i.e., closed under multiplication
  and adjoints) and 2. that self-adjoint (i.e. ``real valued'') elements of $\scr{A}$ represent the
  (projection valued) observables of the system in question.} of the system. The simple rule we have to
follow is: states are described by density operators in $\scr{A}$.  

This point of view is very useful if we consider \emph{channels} which transform one kind of information
into another, e.g. quantum to classical or hybrid. They are most naturally described in terms of
\emph{completely positive}, \emph{trace preserving} maps $T:\scr{A} \to \scr{B}$, where $\scr{A}$ and
$\scr{B}$ denote the observable algebras of the input respectively output systems, and $T(\rho)$ is the state
at the output side of the channel if the input system was in the state $\rho$. Alternatively we can
consider the \emph{dual} $T^*: \scr{B} \to \scr{A}$ of $T$, which is characterized by the condition
$\tr(T^*(A)\rho) = \tr(AT(\rho))$. It describes the operation in the Heisenberg picture, while $T$ is the
Schr{\"o}dinger picture representation. 
The following list summarizes some special cases (arising from different choices for $\scr{A}$ and
$\scr{B}$) which will be relevant for the rest of the paper.

\begin{itemize}
\item 
\hd{Quantum operations:} If $\scr{A} = \scr{B} = \scr{B}(\scr{H})$ the map $T$ describes a quantum
  operation. The most simple case is just unitary time-evolution, i.e. $T(\rho) = U\rho U^*$ with unitary
  operator $U$. In general however, we have to take interactions with additional, unobservable degrees of
  freedom into account (``environment'') and $T$ becomes
  \begin{equation} \label{eq:3}
    T(\rho) = \tr_\scr{K} \bigl(U (\rho \otimes \rho_0) U^*\bigr),
  \end{equation}
  where $\scr{K}$ and $\rho_0$ denote Hilbert space and initial state of the environment (which can be
  chosen to be pure) and $U$ describes now the common evolution of both systems. It is a simple
  consequence of Stinespring's theorem \cite{StSpr} that each quantum operation can be written this way. 
\item 
\hd{Observables:} If $\scr{A}$ is quantum ($\scr{A}=\scr{B}(\scr{H})$) and $\scr{B}$ is classical
  ($\scr{B} = \scr{C}(X)$) we can define $T^{(x)} = T^*(\kb{x})$. It is easy to see that the family
  $T^{(x)} \in \scr{B}(\scr{H})$, $x \in X$ of operators forms a POV measure, hence $T$ describes a
  (generalized) observable and $\tr(\rho T^{(x)})$ is the probability to measure the value $x$ on systems in the
  state $\rho \in \scr{B}(\scr{H})$. We will identify in the following the observable $T$ with the family
  $T^{(x)}$ and write: ``let $T = (T^{(1)}, \ldots, T^{(n)})$ be an observable''. Note that this
  interpretation makes sense as well, if we insert for  $\scr{A}$ a classical or hybrid algebra. Hence we
  can look at the corresponding observables as special cases of quantum observables. 
\item 
\hd{Instruments:} 
  If we are interested in the state of the quantum system after the measurement (in addition to the
  measuring result), we have to consider ``\emph{Instruments}'' i.e. channels with quantum input
  ($\scr{A} = \scr{B}(\scr{H})$) and hybrid output ($\scr{B} = \scr{B}(\scr{H}) \otimes \scr{C}(X)$). To each
  instrument $T$ and each $x \in X$ we can associate a (non trace preserving!) quantum operation $T_x:
  \scr{B}(\scr{H}) \to \scr{B}(\scr{H})$ by $T_x^*(A) = T^*(A \otimes \kb{x})$. For each input state $\rho$ the
  density operator $\tr(T_x\rho)^{-1}T(\rho)$ describes the state of the system after the measurement if the
  value $x \in X$ was obtained, while the probability to measure $x \in X$ is given by $\tr(T_x\rho)$. Hence the
  observable (i.e. the POV measure) associated to $T$ is $T^{(x)} = T^*(\Bbb{1} \otimes \kb{x})$, $x \in X$. 
\item 
\hd{Parameter dependent instruments:}  Finally, let us consider a channel with hybrid input and output,
  i.e. $\scr{A} = \scr{B} = \scr{B}(\scr{H}) \otimes \scr{C}(X)$. For each $x \in X$ we get an instrument $T_x$
  by $T_x(\rho) = T(\rho \otimes \kb{x})$. Hence $T$ describes an instrument whose behavior depends on the additional
  classical input data $x \in X$.
\end{itemize}

\section{Coin tossing protocols}
\label{sec:coin-toss-prot}

Two players (as usual called Alice and Bob) are separated from each other and want to create a
random bit, which can take both possible values with equal probability. However they do not trust each
other and there is no trustworthy third person who can flip the coin for them. Hence they only can
exchange data until they have agreed on a value $0$ or $1$ or until one player is convinced that the
other is cheating; in this case we will write $\draw$ for the corresponding outcome.


To describe such a \emph{coin tossing protocol} mathematically, we need three observable algebras
$\Al$, $\Bl$ and $\Pl$, where $\Al$ and $\Bl$ represent \emph{private information}, which is only accessible
by Alice and Bob respectively -- Alice's and Bob's ``notepad'' -- while $\Pl$ is a \emph{public} area,
which is used by both players to exchange data. We will call it in the following the ``mailbox''. Each
of the three algebras $\Al$, $\Bl$ and $\Pl$  contain in general a classical and a quantum part, i.e. we
have $\Al = \scr{C}(X_A) \otimes \scr{B}(\scr{H}_A)$ and similar for $\Bl$ and $\Pl$. A typical choice is
$\scr{H}_A = \scr{H}^{\otimes n}$ and $X_A = \Bbb{B}^m$ where $\scr{H} = \Bbb{C}^2$ and $\Bbb{B}$ denotes the
field with two elements -- in other words Alice's notepad consists in this case of $n$ qubits and
$m$ classical bits.

If Alice wants to send data (classical or quantum) to Bob, she has to store them in the mailbox system,
where Bob can read them off in the next round. Hence each processing step of the protocol (except the
first and the last one) can be described as follows: Alice (or Bob) uses her own private data and the
information provided by Bob (via the mailbox) to perform some calculations. Afterwards she writes the
results in part to her notepad and in part to the mailbox. An operation of this kind can be described by
a completely positive map $T_A: \Al \otimes \Pl \to \Al \otimes \Pl$, or (if executed by Bob) by $T_B: \Pl \otimes \Bl \to \Pl
\otimes \Bl$.   

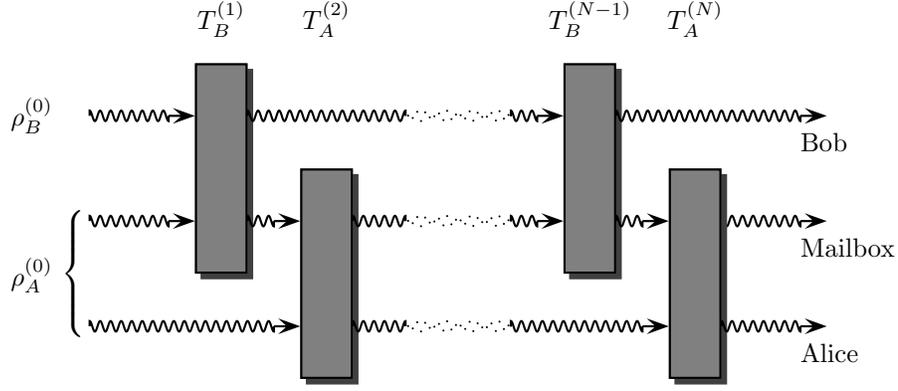
\begin{figure}[t]
  \begin{center}
    \begin{picture}(12,7)
      \psset{coilarm=0,coilaspect=0,coilheight=1,coilwidth=.2}
      \psframe[fillcolor=meascolor,fillstyle=solid,shadow=true](1,2)(2,6)
      \pscoil{-}(-1,5)(0.5,5)
      \psline{->}(0.5,5)(1,5)
      \pscoil{-}(-1,3)(0.5,3)
      \psline{->}(0.5,3)(1,3)
      \pscoil{-}(2,5)(5,5)
      \pscoil{-}(2,3)(2.5,3)
      \psline{->}(2.5,3)(3,3)
      \psframe[fillcolor=meascolor,fillstyle=solid,shadow=true](3,0)(4,4)
      \pscoil{-}(-1,1)(2.5,1)
      \psline{->}(2.5,1)(3,1)
      \pscoil{-}(4,3)(5,3)
      \pscoil{-}(4,1)(5,1)
      \pscoil[linestyle=dotted]{-}(5,1)(7,1)
      \pscoil[linestyle=dotted]{-}(5,5)(7,5)
      \pscoil[linestyle=dotted]{-}(5,3)(7,3)
      \pscoil{-}(7,1)(9.5,1)
      \pscoil{-}(7,5)(7.5,5)
      \psline{->}(9.5,1)(10,1)
      \psline{->}(7.5,5)(8,5)
      \pscoil{-}(7,3)(7.5,3)
      \psline{->}(7.5,3)(8,3)
      \psframe[fillcolor=meascolor,fillstyle=solid,shadow=true](8,2)(9,6)
      \pscoil{-}(9,3)(9.5,3)
      \psline{->}(9.5,3)(10,3)
      \pscoil{-}(11,3)(12.5,3)
      \psline{->}(12.5,3)(13,3)
      \pscoil{-}(9,5)(12.5,5)
      \psline{->}(12.5,5)(13,5)
      \psframe[fillcolor=meascolor,fillstyle=solid,shadow=true](10,0)(11,4)
      \pscoil{-}(11,1)(12.5,1)
      \psline{->}(12.5,1)(13,1)
      \rput[b](1.5,6.5){$T_B^{(1)}$}
      \rput[b](8.5,6.5){$T_B^{(N-1)}$}
      \rput[b](3.5,6.5){$T_A^{(2)}$}
      \rput[b](10.5,6.5){$T_A^{(N)}$}
      \rput[l](-2.5,5){$\rho_B^{(0)}$}
      \rput[l](-2.5,2){$\rho_A^{(0)}$ \LARGE $\Bigg\{$}
      \rput[l](12.5,0.5){Alice}
      \rput[l](12.5,2.5){Mailbox}
      \rput[l](12.5,4.5){Bob}
    \end{picture}
    \caption{Schematic picture of a quantum coin-tossing protocol. The curly arrows stands for the flow of
      quantum or classical information or both.}
    \label{fig:schema}
  \end{center}
\end{figure}

Based on these structures we can describe a coin tossing protocol as follows: At the beginning Alice and
Bob prepare their private systems in some initial state. Alice uses in addition the mailbox system to
share some information about her preparation with Bob, i.e. Alice prepares the system $\Al \otimes \Pl$ in a
(possibly entangled, or at least correlated) state $\rho_A^{(0)}$, while Bob prepares his notepad in the state
$\rho_B^{(0)}$. Hence the state of the composite system becomes $\rho^{(0)} = \rho_A^{(0)} \otimes \rho_B^{(0)}$. Now Alice
and Bob start to operate alternately\footnote{This means we are considering only turn based protocols. If
  special relativity, and therefore finite propagation speed for information, is taken into account it
  can be reasonable to consider simultaneous exchange of information; cf. e.g. \cite{SRQCT} for details.}
on the system, as described in the last paragraph, i.e. Alice in terms of operations $T_A: \Al \otimes \Pl \to
\Al \otimes \Pl$ and Bob with $T_B : \Pl \otimes \Bl \to \Pl \otimes \Bl$. After $N$ rounds\footnote{Basically $N$ is the
  maximal number of rounds: After $K < N$ steps Alice (Bob) can apply identity maps, i.e. $T_A^{(j)} =
  \Id$ for $j > K$.} the systems ends therefore in the state (cf. Figure \ref{fig:schema})
\begin{equation} \label{eq:1}
  \rho^{(N)} = (T_A^{(N)} \otimes \Id_B)(\Id_A \otimes T_B^{(N-1)})  \cdots (T_A^{(2)} \otimes \Id_B)(\Id_A \otimes T_B^{(1)}) \rho^{(0)},
\end{equation}
where $\Id_A$, $\Id_B$ are the identity maps on $\Al$ and $\Bl$. Note that we have assumed here without
loss of generality that Alice performs the first (i.e. providing the initial preparation of the mailbox)
and the last step (applying the operation $T_A$). It is obvious how we have to change the following
discussion if Bob starts the game or if $N$ is odd. To determine the result Alice and Bob perform
measurements on their notepads. The corresponding observables $E_A = (E_A^{(0)}, E_A^{(1)},
E_A^{(\draw)})$ and $E_B = (E_B^{(0)}, E_B^{(1)}, E_B^{(\draw)})$ can have the three possible outcomes $X
= \{0,1,\draw\}$, which we  have described already above. The tuples  
\begin{equation}
  \sigma_A = (\rho_A^{(0)}; T_A^{(2)}, \ldots, T_A^{(N+2)}; E_A), \quad \sigma_B = (\rho_B^{(0)}; T_B^{(1)}, \ldots, T_B^{(N+1)}; E_B)
\end{equation}
consists of all parts of the protocol Alice respectively Bob can influence. Hence we will call $\sigma_A$
Alice's and $\sigma_B$ Bob's strategy. The sets of all strategies of Alice respectively Bob are denoted by
$\Sigma_A$ and $\Sigma_B$. Note that $\Sigma_A$ depends only on the algebras $\Al$ and $\Pl$ while $\Sigma_B$ depends on
$\Bl$ and $\Pl$.  Occasionally it is useful to emphasize this dependency (the number of rounds is kept
fixed in this paper). In this case we write $\Sigma_A(\Al,\Pl)$ and $\Sigma_B(\Bl,\Pl)$ instead of $\Sigma_A$ and
$\Sigma_B$. The probability that Alice gets the result $a\in X$ if she applies the strategy $\sigma_A \in \Sigma_A$ and Bob
gets $b \in X$ with strategy $\sigma_B \in \Sigma_B$ is 
\begin{equation}
  \Bbb{P}(\sigma_A,\sigma_B;a,b) = \tr\bigl[(E_A^{(a)} \otimes \Bbb{1} \otimes E_B^{(b)}) \rho^{(N)}\bigr].
\end{equation}

If both measurements in the last step yield the same result $a=b=0$ or $1$ the procedure is successful (and
the outcome is $a$). If the results differ or if one player signals $\draw$ the protocol fails. As stated
above we are interested in protocols which do not fail and which produce $0$ and $1$ with equal
probability. Another crucial requirement concerns \emph{security}: Neither Alice nor Bob should be able to
improve the probabilities of the outcomes $0$ or $1$ by ``cheating'', i.e. selecting strategies which
deviate from the predefined protocol. At this point it is crucial to emphasize that we do not make any
restricting assumptions about the resources Alice and Bob can use to cheat -- they are potentially
unlimited. This includes in particular the possibility of arbitrarily large notepads. In the next
definition this is expressed by the (arbitrary) algebra $\scr{R}$.

\begin{defi} \label{def:2}
  A pair of strategies $(\sigma_A, \sigma_B) \in \Sigma_A(\Al,\Pl) \times \Sigma_B(\Bl,\Pl)$ is called a (strong) \emph{coin tossing
    protocol} with bias $\epsilon \in [0,1/2]$ if the following conditions holds for any (finite dimensional)
  observable algebra $\scr{R}$ 
  \begin{enumerate}
  \item \label{item:6}
    \emph{Correctness:} $\Bbb{P}(\sigma_A, \sigma_B; 0,0)=\Bbb{P}(\sigma_A, \sigma_B; 1,1) = \frac{1}{2}$,
  \item  \label{item:4}
    \emph{Security against Alice:} $\forall \sigma_A' \in \Sigma_A(\scr{R} ,\Pl)$ and $\forall x \in \{0,1\}$ we have
    \begin{equation}
      \Bbb P(\sigma_A', \sigma_B; x, x) \leq \frac{1}{2}+\epsilon
    \end{equation}
  \item \label{item:5}
    \emph{Security against Bob:} $\forall \sigma_B' \in \Sigma_B(\scr{R}, \Pl)$ and $\forall x \in \{0,1\}$ we have 
    \begin{equation}
      \Bbb P(\sigma_A, \sigma_B'; x, x) \leq \frac{1}{2}+\epsilon
    \end{equation}
  \end{enumerate}
\end{defi}

The two security conditions in this definition imply that neither Alice nor Bob can increase the
probability of the outcome $0$ \emph{or} $1$ beyond the bound $1/2+\epsilon$. However it is more natural to
think of coin tossing as a game with payoff defined according to the following table
\begin{equation} \label{eq:6}
  \mbox{\begin{tabular}{|l|c|c|}
    \hline 
    & Alice &  Bob \\ \hline \hline
    a=b=0 & 1 & 0 \\ \hline
    a=b=1 & 0 & 1 \\ \hline
    other & 0 & 0 \\ \hline
  \end{tabular} }
\end{equation}
This implies that Alice tries to increase only the probability for the outcome $0$ and not for $1$ while
Bob tries to do the contrary, i.e. increase the probability for $1$. This motivates the following
definition. 

\begin{defi} \label{def:3}
  A pair of strategies $(\sigma_A,\sigma_B) \in \Sigma_A(\Al,\Pl) \times \Sigma_B(\Bl,\Pl) $ is called a \emph{weak coin tossing
    protocol}, if item \ref{item:6} of Definition \ref{def:2} holds, and if items \ref{item:4} and
  \ref{item:5} are replaced by 
  \begin{enumerate}
  \item[\ref{item:4}']
    \emph{Weak security against Alice:} $\forall \sigma_A' \in \Sigma_A(\scr{R} ,\Pl)$ we have 
    \begin{equation}
      \Bbb P(\sigma_A', \sigma_B; 0, 0) \leq \frac{1}{2}+\epsilon, 
    \end{equation}
  \item[\ref{item:5}']
    \emph{Weak security against Bob:} $\forall \sigma_B' \in \Sigma_B(\scr{R} ,\Pl)$ we have 
    \begin{equation}
          \Bbb P(\sigma_A, \sigma_B'; 1, 1) \leq \frac{1}{2}+\epsilon. 
    \end{equation}
  \end{enumerate}
  Here $\scr{R}$ stands again for any finite dimensional (but arbitrarily large) observable algebra. 
\end{defi}

Good coin tossing protocols are of course those with a small bias. Hence the central question is: What
is the smallest bias which we have to take into account, and how do the corresponding optimal strategies
look like? To get an answer, however, is quite difficult. Up to now there are only partial results
available (cf. Section \ref{sec:bounds} for a summary). 

Other but related questions arises if we exploit the game theoretic nature of the problem. In this
context it is reasonable to look at a whole class of quantum games, which arises from the scheme developed
up to now. We only have to fix the algebras\footnote{In contrast to the security definitions given above
  this means that we assume limited recourses (notepads) of Alice and Bob. This simplifies the analysis
  of the problem and should not be a big restriction (from the practical point of view) if the notepads
  are fixed but very large.} $\Al$, $\Bl$ and $\Pl$ and to specify a payoff matrix as in Equation
(\ref{eq:6}). The latter, however, has to be done carefully. If we consider instead of (\ref{eq:6}) the
payoff 
\begin{equation}
  \mbox{\begin{tabular}{|l|c|c|}
    \hline 
    & Alice &  Bob \\ \hline \hline
    a=b=0 & 1 & -1 \\ \hline
    a=b=1 & -1 & 1 \\ \hline
    other & 0 & 0 \\ \hline
  \end{tabular} }
\end{equation}
we get a zero sum game, which seems at a first look very reasonable. Unfortunately it admits a very simple
(and boring) optimal strategy: Bob produces always the outcome $1$ on his side while Alice claims always
that she has measured $0$. Hence they never agree and nobody has to pay. The game from Equation
(\ref{eq:6}) does not suffer from this problem, because a draw is for Alice as bad as the case $a=b=1$
where Bob wins.  

\section{Classical coin tossing}
\label{sec:class-coin-toss}

Let us now add some short remarks on classical coin tossing, which is included in the general scheme just
developed as a special case: We only have to choose classical algebras for $\Al$, $\Bl$ and $\Pl$, 
i.e. $\Al = \scr{C}(X_A)$, $\Bl = \scr{C}(X_B)$ and $\Pl = \scr{C}(X_M)$. 
The completely  positive maps $T_A$ and $T_B$ describing the operations performed by Alice and Bob are in
this case given by matrices of \emph{transition probabilities} (see Sect. 3.2.3 of \cite{PRepQI} to see
how to relate these matrices to the operations $T$). This implies in particular that the strategies in
$\Sigma_A$, $\Sigma_B$ are in general \emph{mixed strategies}. 
This is natural -- there is of course no classical coin tossing protocol consisting of pure strategies,
because it would lead always to the same result (either always $0$ or always $1$). However, we can
decompose each mixed strategy in a unique way into a convex linear combination of pure strategies, and
this can be used to show that there is no classical coin tossing protocol, which admits the kind of
security contained in Definition \ref{def:2} and \ref{def:3}. 

\begin{prop}
  There is no (weak) classical coin tossing protocol with bias $\epsilon < \frac{1}{2}$.
\end{prop}

\begin{proof}
  Assume a classical coin tossing protocol $(\sigma_A,\sigma_B)$ is given. Since its outcome is by definition
  probabilistic, $\sigma_A$ or $\sigma_B$ (or both) are mixed strategies which can be decomposed (in a unique way)
  into pure strategies. Let us denote the sets of pure strategies appearing in this decomposition by
  $\Sigma_A'$, $\Sigma_B'$. Since the protocol $(\sigma_A,\sigma_B)$ is correct, each pair $(s_A,s_B) \in \Sigma_A'\times\Sigma_B'$ leads to a
  valid outcome, i.e. either $0$ or $1$ on both sides. Hence there are two possibilities to construct a 
  zero-sum game, either Alice wins if the outcome is $0$ and Bob if it is $1$ or the other way round. In
  both cases we get a zero-sum two-person game with perfect information, no chance moves\footnote{That
    means there are no \emph{outside} probability experiments like dice   throws.} and only two
  outcomes. In those games one player has a \emph{winning strategy} (cf. Sect. 15.6, 15.7 of
  \cite{gametheory}),  i.e. if she (or he) follows that strategy she wins with certainty, no matter
  which strategy the opponent uses. This includes in particular the case where the other player is honest
  and follows the protocol. If we apply this arguments to both variants of the game, we see that either
  one player could force both possible outcomes or one bit could be forced by both players. Both cases
  only fit into the definition of (weak) coin tossing if the bias is $1/2$. This proves the proposition.    
\end{proof}

Note that the proof is not applicable in the quantum case (in fact there are coin tossing protocols with
bias less than $1/2$ as we will see in Section \ref{sec:particular-example}). One reason is that in the
quantum case one does not have perfect information. E.g. if Alice sends a qubit to Bob, he does not know
what qubit he got. He could perform a measurement, but if he measures in a wrong basis, he will
inevitably change the qubit.  

Another way to circumvent the negative result of the previous proposition is to weaken the assumption
that both players can perform \emph{any} operation on their data. A possible practical restriction which
come into mind immediately is limited computational power, i.e. we can assume that no player is able to
solve intractable problems like factorization of large integers in an acceptable time. Within the definition
given above this means that Alice and Bob do not have access to all strategies in $\Sigma_A$ and $\Sigma_B$ but
only to certain subsets. Of course, such additional restrictions can be imposed as well in the quantum
case. To distinguish the much stronger security requirements in Definition \ref{def:2} and \ref{def:3} a
protocol is sometimes called \emph{unconditionally secure}, if no additional assumptions about the
accessible cheating strategies are necessary (loosely speaking: the ``laws of quantum mechanics'' are the
only restriction). 

\section{The unitary normal form}
\label{sec:unitary-normal-form}

A special class of quantum coin tossing arises if: 1. all algebras are quantum, i.e. $\Al =
\scr{B}(\scr{H}_A)$, $\Bl = \scr{B}(\scr{H}_B)$ and $\Pl = \scr{B}(\scr{H}_M)$ with Hilbert spaces
$\scr{H}_A$, $\scr{H}_B$ and $\scr{H}_M$; 2. the initial preparation is pure: $\rho_a = \kb{\psi_A}$ and
$\kb{\psi_B}$ with $\psi_A \in \scr{H}_A \otimes \scr{H}_M$ and $\psi_B \in \scr{H}_B$; 3. the operations $T_A^{(j)}$,
$T_B^{(k)}$ are unitarily implemented: $T_A^{(j)}(\rho) = U_A^{(j)} \rho U_A^{(j)*}$ with a unitary operator
$U_A^{(j)}$ on $\scr{H}_A \otimes \scr{H}_M$ and something similar holds for Bob and 4. the observables $E_A$,
$E_B$ are projection valued. It is easy to see that the corresponding strategies $(\sigma_A,\sigma_B) \in \Sigma_A \times \Sigma_B$
do not admit a proper convex decomposition into other strategies. Hence we will call them in the
following \emph{pure strategies}. In contrast to the classical case it is possible to construct correct
coin tossing protocols with pure strategies. The following proposition was stated for the first time (in
a less explicit way) in \cite{Mayers97} and shows that we can replace a mixed strategy always by a pure
one without loosing security. 

\begin{prop} \label{normalform}
  For each strategy $\sigma_A \in \Sigma_A(\Al,\Pl)$ with $\Al \subset \scr{B}(\scr{H}_A)$ there is a Hilbert space
  $\scr{K}_A$ and a unitary strategy $\tilde{\sigma}_A \in \Sigma_A(\tilde{\Al},\Pl)$ with $\Al = \scr{B}(\scr{H}_A \otimes
  \scr{K}_A)$ such that 
  \begin{equation} \label{eq:4}
    \Bbb{P}(\sigma_A,\sigma_B;x,y) = \Bbb{P}(\sigma_A',\sigma_B;x,y)
  \end{equation}
  holds for all $\sigma_B \in \Sigma_B(\Bl,\Pl)$ (with arbitrary Bob algebra $\Bl$) and all $x,y \in \{0,1,\draw\}$. A
  similar statement holds for Bob's strategies.
\end{prop} 

\begin{proof}
  We will only give a sketch of the proof here; the details are given in \cite{Doe02}. Note first
  that all observable algebras $\Al$, $\Bl$ and $\Pl$ are linear subspaces of pure quantum algebras,
  i.e. $\Al \subset \scr{B}(\scr{H}_A)$, $\Bl \subset \scr{B}(\scr{H}_B)$ and $\Pl \subset \scr{B}(\scr{H}_M)$. In addition
  it can be shown that Alice's operations $T_A: \Al \otimes \Pl \to \scr{B}(\scr{H}_A) \otimes \scr{B}(\scr{H}_M)$ can
  be extended to a channel $\tilde{T}_A: \scr{B}(\scr{H}_A) \otimes \scr{B}(\scr{H}_M) \to \scr{B}(\scr{H}_A) \otimes
  \scr{B}(\scr{H}_M)$, i.e. a quantum operation \cite{Paulsen}; something similar holds for Bob's
  operations. Hence we can restrict the proof to the case where all three observable algebras are
  quantum. Now the statement basically follows from the fact that we can find for each item in the
  sequence $T_A = (\rho_A; T_A^{(2)}, \ldots, T_A^{(N)}; E_A)$ a ``dilation''. For the operations $T_A^{(j)}$
  this is just the ancilla representation given in Equation (\ref{eq:3}), i.e.
  \begin{equation}
    T_A^{(j)}(\rho) = \tr_2 \bigl( V^{(j)} (\rho \otimes \kb{\phi^{(j)}}) V^{(j)*}\bigr) 
  \end{equation}
  with a Hilbert space $\scr{L}^{(j)}$, a unitary $V^{(j)}$ on $\scr{H}_A \otimes \scr{L}^{(j)}$ and a pure
  state $\phi^{(j)} \in \scr{L}^{(j)}$ (and $\tr_2$ denotes the partial trace over
  $\scr{L}^{(j)}$). Similarly, there is a Hilbert space $\scr{L}^{(0)}$ and a pure state $\phi^{(0)} \in
  \scr{H}_A \otimes \scr{L}^{(0)}$ such that
  \begin{equation}
    \rho_A = \tr_2(\kb{\phi^{(0)}})
  \end{equation}
  holds (i.e. $\phi^{(0)}$ is the purification of $\rho_A$; cf. \cite{PRepQI} Sect. 2.2), and finally we have a
  Hilbert space $\scr{L}^{(N+2)}$, a pure state $\phi^{(N+2)}$ and a projection valued measure
  $F^{(0)}, F^{(1)}, F^{(\draw)} \in \scr{B}(\scr{H}_A \otimes \scr{L}^{(N+2)})$ with
  \begin{equation}
    \tr(E_A^{(x)}\rho) = \tr\bigl(F^{(x)} (\rho \otimes \kb{\phi^{(N+2)}})\bigr),
  \end{equation}
  this is another concequence of Stinesprings theorem. Now we can define the unitary strategy
  $\tilde{\sigma}_A$ as follows: 
  \begin{gather}
    \scr{K}_A = \scr{L}^{(0)} \otimes \scr{L}^{(2)} \otimes \ldots \otimes \scr{L}^{(N)} \otimes \scr{L}^{(N+2)}\\
    \psi_A = \phi^{(0)} \otimes \phi^{(2)} \otimes \cdots \otimes \phi^{(N)} \otimes \phi^{(N + 2)}\\
    U_A^{(j)} = \Bbb{1}_0 \otimes \Bbb{1}_2 \otimes \cdots \otimes V^{(j)} \otimes \cdots \otimes \Bbb{1}_{N} \otimes \Bbb{1}_{N+2} \label{eq:5}  \\
    \tilde{E}_A^{(x)} = \Bbb{1}_0 \otimes \cdots \otimes \Bbb{1}_{N} \otimes F^{(x)}, 
  \end{gather}
  where $\Bbb{1}_k$ denotes the unit operator on $\scr{L}^{(k)}$ and in Equation (\ref{eq:5}) we have
  implicitly used the canonical isomorphism between $\scr{H}_A \otimes \scr{K}_A$ and $\scr{L}^{(0)} \otimes \cdots \otimes
  \scr{H}_A \otimes \scr{L}^{(j)} \otimes \ldots \otimes \scr{L}^{(N+2)}$ . What remains to show, but is omitted
  here, is to check that this $\tilde{\sigma}_A$ satisfies Equation (\ref{eq:4}).
\end{proof} 

This result allows us to restrict many discussions to pure strategies. This is very useful for
the proof of no-go theorems or for calculations of general bounds on the bias of coin-tossing
protocols. This concerns in particular the results in \cite{Ambainis} which apply, due to Proposition
\ref{normalform} immediately to the general case introduced in Section \ref{sec:coin-toss-prot}. Many
concrete examples (cf. the next section) are however mixed protocols and to rewrite them in a pure form
is not necessarily helpful.  

\section{A particular example}
\label{sec:particular-example}

In this section we are giving a concrete example for a  strong coin tossing protocol. It has a bias of
$\epsilon=0.25$ and is derived from a quantum bit commitment protocol. (a procedure related to coin tossing)
given in \cite{QBC}.  Bit commitment is another two person protocol which is related to coin tossing. It is
always possible to construct a coin tossing protocol from a bit commitment protocol but not the other way
round (cf. \cite{SRQCT}). Hence statements about the security of certain bit commitment protocols can be 
translated into statements about the bias of the related coin tossing protocols. This shows together with
\cite{QBC} that the given protocol has the claimed bias. 

\subsection{The protocol}

In this protocol we take $\scr H _A=\scr H_M = \Bbb C^3, \scr H_B = \Bbb C^3 \otimes \Bbb C^3$ plus classical
parts of at most 2 bits for each notepad. The canonical base in the Hilbert space $\Bbb C^3$ is denoted
by $\ket{i}, i =0,1,2$  

\begin{enumerate}
\item{\emph{preparation step:}} Alice throws a coin, the result is $b_A \in \{0,1\}$, with probability $1/2$
  each. She stores the result and prepares the system $\scr B(\scr H_A) \otimes \scr B(\scr H_M)$ in the state
  $\kb{\psi_{b_A}}$, where $\ket{\psi_0}=\frac{1}{\sqrt{2}}\left(\ket{0,0}+\ket{1,2}\right)$ and
  $\ket{\psi_1}=\frac{1}{\sqrt{2}}\left(\ket{1,1}+\ket{0,2}\right)$ are orthogonal to each other. Bob throws
  a similar coin, and stores the result $b_B$. The initial preparation of his quantum part is arbitrary. 

\item \label{cheatB} Bob reads the mailbox (i.e. swaps it with the second part of his Hilbert space)
  and sends $b_B$ to Alice. 

\item \label{cheatA} Alice receives $b_B$ and puts her remaining quantum system into the mailbox.  

\item Bob reads the mailbox and puts the system into the first slot of this quantum register. 

\item{\emph{results:}} The result on Alice's side is $b_A\oplus b_B$, where $\oplus$ is the addition modulo 2. Bob
  performs a projective measurement on his quantum system with $P^{(0)}=\kb{\psi_0},P^{(1)}=\kb{\psi_1}$ and
  $P^{(\draw)}=\Bbb 1 - P^{(0)}-P^{(1)}$, with result $b_A^\prime$. If everybody followed the protocol
  $b_A^\prime=b_A$. So the result on Bob's side is $b_A^\prime\oplus b_B$. 
\end{enumerate}

\subsection{Possible cheating strategies}

Now we will give possible cheating strategies for each party which lead to the maximal probability of
achieving the preferred outcome. For simplicity we just look at the case where Alice prefers the
outcome to be 0, whereas Bob prefers it to be 1, cheating strategies for the other cases are easily
derivable. A cheating strategy for Bob is to try to distinguish in 
step \ref{cheatB} whether Alice has prepared $\ket{\psi_0}$ or $\ket{\psi_1}$. For this purpose he performs
the measurement $(\kb{0},\kb{1},\kb{2})$. If the result $c_B \neq 2$ (the probability for this in either
case is $1/2$) he can identify $b_A=c_B$ and set $b_B=c_B\oplus 1$ to achieve the overall result $1$. If
$c_B=2$ holds, he has not learned anything about $b_A$. In that case he just continues with the protocol
and hopes for the desired result, which appears with the probability $1/2$.\footnote{After that measurement
  he is no longer able to figure out which outcome occurs on Alice's side, so he just sets his outcome
  to 1. A similar situation occurs in the cheating strategy for Alice, but she is in neither case able to
  predict the outcome on Bob's side with certainty.} So the total probability for Bob to achieve the
result 0 is $\frac{1}{2}+\frac{1}{2} \cdot \frac{1}{2} =\frac{3}{4}$. 

A cheating strategy for Alice is to set in the initial step $b_A=0$ and to prepare the system $\scr
B(\scr H_A) \otimes \scr B(\scr H_M)$ in the state
$\ket{\tilde{\psi_0}}=\frac{1}{\sqrt{6}}\left(\ket{0,0}+\ket{0,1}+2 \cdot \ket{1,2}\right)$. Then she continues 
until step \ref{cheatA}. If $b_B=0$ she just continues with the protocol. Then the probability that in
the last step Bob measures $b_B^\prime=0$ equals $\tr(\kb{\tilde{\psi_0}}\cdot
\kb{\psi_0})=|\BK{\psi_0}{\tilde{\psi_0}}|^2=\frac{3}{4}$. If $b_B=1$ she first applies a unitary operator, which
swaps $\ket 0$ and $\ket 1$, on her system before she sends it to Bob. The state on Bob's side is than
$\kb{\tilde{\psi_1}}$ with $\ket{\tilde{\psi_1}}=\frac{1}{\sqrt{6}}\left(\ket{1,0}+\ket{1,1}+2 \cdot
  \ket{0,2}\right)$. The probability that Bob measures $b_B^\prime=1$ equals $\tr(\kb{\tilde{\psi_1}}\cdot
\kb{\psi_1})=|\BK{\psi_1}{\tilde{\psi_1}}|^2=\frac{3}{4}$. So the total probability for Alice to get the outcome 0
is $\frac{1}{2} \cdot \frac{3}{4} + \frac{1}{2} \cdot \frac{3}{4}=\frac{3}{4}$.

\section{Conclusions}
\label{sec:bounds}

The previous example shows that quantum coin tossig admits, in contrast to the classical case, a
nontrivial bias. However, how secure quantum coin tossing really is? Can we reach the optimal case ($\epsilon =
0$)? The answer actually is ``no'', or to state it more explicitly:

\begin{thm}
There is no (strong or weak) coin tossing protocol with bias $\epsilon=0$.
\end{thm}

This was first proven by Mayers, Salvail and Chiba-Kohno \cite{MaSaChi}. Later on Ambainis recalls the
arguments in a more explicit form 
\cite{Ambainis}\footnote{The first attempt for a proof, given by Lo and Chau \cite{LoCh98}. However,
  its validity  is restricted to the case where `cheating' always influences the probabilities of
  \emph{both} valid outcomes. More precisely they demand that the probabilities for the outcomes $0$ and 
  $1$ are equal, for any cheating strategy. This restriction is too strong, even if Alice and Bob
  sit together and throw a real coin one of them can always say he (she) does not accept the
  result (and for example refuses to pay his loss) and so put the probability for one outcome to zero
  while the probability for the other one and the outcome invalid are $1/2$ each.}. 
It is still an open question, whether there exists quantum coin tossing protocols with bias arbitrarily
near to zero. Ambainis also shows that a coin tossing protocol with a bias of at most $\epsilon$ must use at
least $\Omega (\log \log \frac{1}{\epsilon})$ rounds of communication. Although in that paper he gives only the proof
for strong coin tossing, it holds in the weak case as well.  It follows that a protocol cannot be made
arbitrarily secure  (i.e. have a sequence of protocols with $\epsilon \to 0$) with just increasing the amount of
information exchanged in each step. The number of rounds has to go to infinity (although very slow). 

The strong coin tossing protocol given in section \ref{sec:particular-example} has a bias of
$\epsilon=0.25$. Another one with the same bias is given by Ambainis \cite{Ambainis}. No strong protocol with
provable smaller bias is known yet. The best known weak protocol is given by Spekkens and Rudolph
\cite{WQCT} and has a bias of $\epsilon=\frac{1}{\sqrt{2}}-\frac{1}{2}=0.207\ldots$\ . Although this is still far
from arbitrarily secure, it shows another distinction between classical and quantum information, as in a
classical world no protocol with bias smaller than $0.5$ is possible. 

Another interesting topic in quantum coin tossing is the question of cheat-sensitivity, that means how
much can each player increase the probability of one outcome without risking being caught cheating. For
more about this cf e.g. \cite{WQCT} or \cite{cheatsensitivity}. 


\end{document}